\documentclass[twocolumn,superscriptaddress,showpacs,aps,amsmath,amssymb]{revtex4}
\usepackage{epsfig}
\usepackage{amsbsy}
\usepackage{amsmath}
\usepackage{amsfonts}
\usepackage{graphicx}
\usepackage{latexsym}

\begin{document}

\title{Single-electron turnstile pumping with high frequencies}
\author{Chuan-Yu Lin}
\affiliation{Department of Physics and Center for Quantum
information Science, National Cheng Kung University, Tainan 70101,
Taiwan}
\author{Wei-Min Zhang}
\email{wzhang@mail.ncku.edu.tw} \affiliation{Department of Physics
and Center for Quantum information Science, National Cheng Kung
University, Tainan 70101, Taiwan}

\begin{abstract}
In this Letter, we present a theoretical analysis to single-electron
pumping operation in a large range of driving frequencies through
the time-dependent tunneling barriers controlled by external gate
voltages. We show that the single-electron turnstile works at the
frequency lower than the characteristic frequency which is
determined by the mean average electron tunneling rate. When the
driving frequency is greater than the characteristic frequency of
electron tunnelings, fractional electron pumping occurs as an effect
of quantum coherence tunneling.
\end{abstract}

\date{June 26,~2011; revised July 25, 2011}


 \maketitle

Single-electron pumps and turnstiles are nanoscale tunneling devices
utilizing controllable transfer of electrons one-by-one synchronized
with alternating external gate voltages. These devices are supposed
to have important applications as current standards and also as
high-frequency amplifiers/detectors in solid-state quantum
computing. Electrons pumping operations have been experimentally
demonstrated with various nanoscale tunneling structures, in
particular, through switching-on and off the tunneling barriers
\cite{setp1,set4,set5,set6,pumps}. These experimental realizations
for single-electron turnstiles are basically operated at relatively
low frequencies ($\sim$ hundreds MHz to a few GHz) with a small
pumped current ($<$ tens  to hundreds pA). To understand the
single-electron pumping mechanism at higher frequencies with a
larger pumping current, we shall closely monitor the electron
transfer in these nanoelectronic devices with a large range of the
signal frequencies. We shall utilize the exact nonequilibrium
quantum transport theory for nanodevices we developed recently
\cite{Jin10083013} and focus on the single-electron device with
time-dependent tunneling barriers controlled by external gate
voltages, to examine the real-time dynamics of the electron transfer
far away from equilibrium.

\begin{figure}
\includegraphics[width=0.95\columnwidth,angle=0]{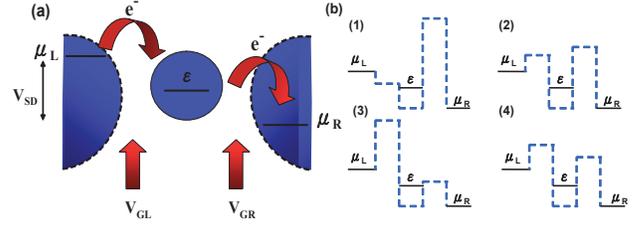}
\caption{(a) A schematic plot of the single-electron turnstiles with
$V_{GL}(t)$ and $V_{GR}(t)$ controlling the tunneling barriers.
(b)The schematic diagrams of the barrier changing at different time
from (1) to (4).} \label{fig1}
\end{figure}

We model the single-electron device by a quantum dot junction of a
single level quantum dot coupling to a source and a drain, as
schematically shown in Fig.~\ref{fig1}(a). The pump or turnstile
operations are realized by imposing the repetitive pulses to tune
the tunneling barriers. As the gate voltages $V_{GL}$ and $V_{GR}$
are applied to gate, the tunneling barriers vary in time. By
modulating the tunnel barriers, it has been observed
\cite{setp1,set4,set5,set6,pumps} that electrons can be transferred
one-by-one between the source and the drain. Here we follow the
experimental setup given in Ref.~\cite{setp1} where the barriers are
modulated in anti-phase and the dot level is kept fixed.
Fig.~\ref{fig1}(b) plots schematically the barrier changing process
from (a) to (d) and then repeat continually in time. Also, a dc bias
voltage $V_{\rm SD}$ is applied between the source and drain to
examine the single-electron pumping dynamics in a large range of
driving frequency.

Based on the recently developed non-equilibrium quantum transport
theory \cite{Jin10083013} which is derived from the exact master
equation for nanoelectronics \cite{Tu08235311}, the transient
electron occupation number in the dot and the transient electron
current flowing from the leads into the dot are given by
\begin{subequations}
\label{curr-char}
\begin{align}
 &n(t) =  v(t,t)+ u(t,t_0)n(t_0)u^{\dag}(t,t_0), \\
 &I_{L,R}(t)= -{2e\over \hbar}{\rm Re}\int_{t_0}^{t} d\tau {\rm
Tr}\Big\{g_{L,R}(t,\tau)v(\tau,t)-\widetilde{g}_{L,R}(t,\tau)\notag
\\ & ~~~~~~ \times u^\dag(t,\tau)
+g_{L,R}(t,\tau)u(\tau,t_0)n(t_0)u^{\dag}(t, t_0)\Big\},
\end{align}
\end{subequations}
respectively, where $L,R$ denote the left and right leads (source
and drain).  The functions $u(\tau, t_0)$ and $v(\tau, t)$ in
Eq.~(\ref{curr-char}) are related to the retarded and correlation
Green functions that satisfy the dissipation-fluctuation
integrodifferential equations of motion
\cite{Tu08235311,Jin10083013}:
\begin{subequations}
\label{uv-eq}
\begin{align}
&\dot{u}(\tau, t_0) +i \varepsilon u (\tau, t_0)+\int_{t_0}^{\tau }
d\tau' g
(\tau ,\tau') u (\tau', t_0) =0 ,  \label{u-e} \\
&v(\tau, t)=\int_{t_0}^{\tau} d\tau' \int_{t_0}^{t} d\tau'' u (\tau,
\tau') \widetilde{g}(\tau', \tau'') u^{\dag}(t,\tau'') , \label{v-e}
\end{align}
\end{subequations}
with the initial condition $u(t_0,t_0)=1$. $n(t_0)$ in
Eq.~(\ref{curr-char}) is the initial electron occupation in the dot.
Here, the non-local time-correlation functions $g (\tau ,\tau')=g_L
(\tau ,\tau')+g_R (\tau ,\tau')$ and $ \widetilde{g} (\tau ,\tau')=
\widetilde{g}_L (\tau ,\tau')+\widetilde{g}_R (\tau ,\tau')$ with
\begin{subequations}
\label{correlation}
\begin{align}
g_{L,R}( \tau ,\tau') &=\int_{-\infty}^{\infty }\frac{d\omega}{2\pi}
J_{L,R}\left(
\omega \right) e^{-i\omega ( \tau -\tau')} ,   \\
\widetilde{g}_{L,R}(\tau ,\tau') &=\int_{-\infty}^{\infty
}\frac{d\omega}{2\pi} J_{L,R}( \omega) f_{L,R}(\omega)e^{-i\omega (
\tau -\tau')} ,
\end{align}
\end{subequations}
in which $f_{L,R}(\omega) =\frac{1}{e^{\beta (\omega-\mu_{L,R})}
+1}$ are the initial electron distribution functions in the leads at
the initial temperature $\beta=1/k_BT$, and  $\mu_{L,R}$ the
corresponding chemical potentials. $J_{L,R}(\omega)=2\pi
\varrho_{L,R}(\omega)|V_{L,R}(\omega)|^2$ are the spectral densities
with $\varrho_{L,R}(\omega)$ being the densities of states of the
leads, and $V_{L,R}(\omega)$ the lead-dot coupling coefficients. The
integration in Eq.~(\ref{uv-eq}) with the kernels $g_{L,R}( \tau
,\tau')$ and $\widetilde{g}_{L,R}(\tau ,\tau')$ encompass all the
back-reaction non-Markovian memory effects between the leads and dot
associating with quantum dissipation and fluctuation. These effects
must be fully taken into account for the accuracy of transient
electron transfer.

\begin{figure}
\includegraphics[width=0.85\columnwidth,angle=0]{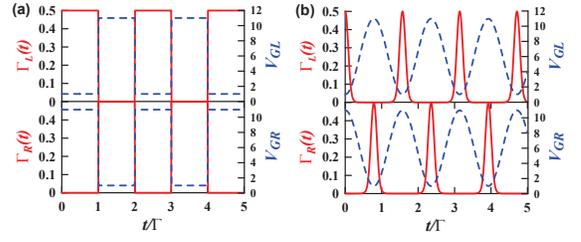}\newline
\caption{The tunneling rates of the left (red solid lines) and right
(blue dashed lines) leads alter with time by (a) a square wave gate
voltage with amplitude $V_{\rm sq}=100\Gamma$ and frequency
$\omega_{\rm sq}=2\pi f_{\rm sq}=\pi\Gamma$ and (b) a sinusoidal
wave modulation with $V_{\sin}=100\Gamma$ and $\omega_{\sin}=2\pi
f_{\sin}=2\Gamma$.} \label{fig2}
\end{figure}

In reality, the spectral densities have more or less a
Lorentzian-type shape \cite{Mac06085324,Tu08235311,Jin10083013},
\begin{align}
J_{L,R}(\omega) = \frac{\Gamma_{L,R}(t) d_{L,R}^{2}}{(\omega
-\mu_{L,R})^{2}+d_{L,R}^{2}},\label{spectral}
\end{align}
where $\Gamma_{L,R}(t)$ are the time-dependent electron tunneling
rate from the leads to the dot that can be controlled by the gate
voltages $V_{GL,GR}(t)$, and $d_{L,R}$ are the bandwidths of the
spectral densities.  The explicit relation between the tunneling
rate and the gate voltage can be obtained by solving the
Schr\"{o}dinger equation of a one-dimensional scattering problem,
\begin{align}
\Gamma_{\alpha}(t)
=\frac{e}{\cosh^2(2k_{\alpha}(t)a)+\varpi_{\alpha}^{2}(t)\sinh^2(2k_{\alpha}(t)a)},\label{rate}
\end{align}
with $k_{\alpha}(t)$
=$\sqrt{\frac{2m^*(V_{G\alpha}(t)-\varepsilon)}{\hbar^{2}}}$ and
$\varpi_{\alpha}(t)$=$\frac{V_{G\alpha}(t)-2\varepsilon}{\sqrt{4\varepsilon(V_{G\alpha}(t)-\varepsilon)}}$,
where $a$ is the width of barrier, and $m^*$ the effective mass of
the electron in the sample. To study electron pumping processes, we
apply a square wave modulation and a sinusoidal wave modulation to
vary the tunneling barriers between the leads and the dot. For the
square wave, $V_{GL}(t)$ and $V_{GR}(t)$ vary from $\varepsilon$ to
$\varepsilon+V_{\rm sq}$ in frequency $f_{\rm sq}$ with a $\pi$
phase shift. For the sinusoidal wave,
$V_{GL}(t)=\varepsilon+V_{\sin}\sin^{2}(2\pi f_{\sin}t)$, and
$V_{GR}(t)=\varepsilon+V_{\sin}\cos^{2}(2 \pi f_{\sin}t)$.
Fig.~\ref{fig2}(a)-(b) plot the corresponding tunneling rate
$\Gamma_{L,R}(t)$.

To monitor the single electron pumping dynamics, a dc bias voltage
$V_{\rm SD}$ is also applied to the leads. The initial temperature
of the leads is taken at $k_BT=0.1 \Gamma$. We also fix the
bandwidth of the spectral density by $d_{L,R}=20 \Gamma$ which is
close to the wide band limit. If we take $\Gamma$=0.1 meV, the
corresponding temperature is about $T \simeq 116$mK. We may also
take the on-site-energy of the dot and the applied bias voltage at
the same scale, e.g. $\varepsilon$=1 meV and $V_{SD}$=2 meV.  All
these parameters are controllable in experiments. To investigate the
turnstile operation at different driving frequency regime, we vary
the driving field from the low to high frequencies. The low and high
frequency regimes are divided with respect to the electron response
time which is determined by the so-called dwell
time\cite{Buttiker96,Guo07}. The dwell time, defined by
$\tau_d$=4$\hbar$/$\Gamma_{L,R}$, is a time scale characterizing the
time of the electron relaxing out the dot. For the above setup of
the system, taking an mean value average tunneling rate
$\overline{\Gamma}_{L,R}$ in each cycle, we obtain $\tau_d$=26.3 ps
for square wave modulation and 63.12 ps for sinusoidal wave
modulation, which correspond to the characteristic frequencies
$f_d=38$ GHz and 15.8 GHz, respectively.

\begin{figure}
\includegraphics[width=1.0\columnwidth,angle=0]{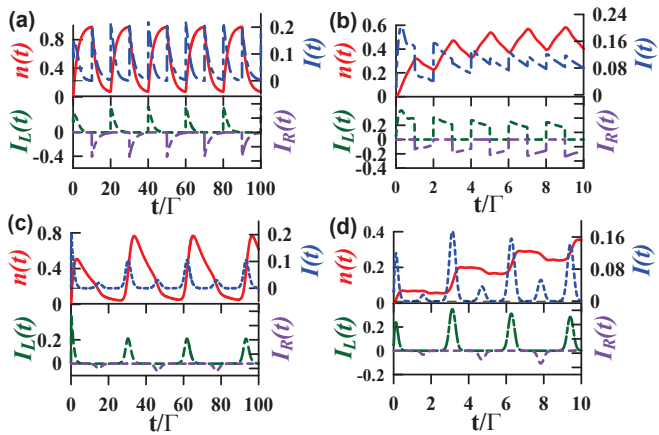}\newline
\caption{The time-dependent electron population (the solid line),
the pumping current (the dash-dotted line), left current (dash-dash
line) and right current (dash-dotted-dotted line) altered by (a)-(b)
the square wave modulation with the driving frequency $f_{\rm
sq}$=7.5GHz and 75GHz, and (c)-(d) the sinusoidal wave modulation
with the driving frequency $f_{\sin}$=5GHz and 50GHz, respectively.}
\label{fig3}
\end{figure}

Now, we discuss how the electron turnstile operates under
time-dependent tunneling barriers. In Fig.~\ref{fig3}(a)-(b), we
plot the time-dependent electron population in the dot, the left and
right current flowing into the dot, given by Eq.~(\ref{curr-char}),
as well as the pumping current $I(t)=\frac{1}{2}[I_L(t)-I_R(t)]$
under the control of a square wave gate voltage. When the left
barrier is opened to the source gate and the right barrier is
closed, the electron tunnels from the source to the dot in each
cycle with an increasing electron population and a positive current,
as shown in Fig.~\ref{fig3}(a). When the left barrier is closed and
the right barrier is opened, the electron tunnels from the dot to
the drain with a decreasing population and a negative current.
However, the device operated under different driving frequency
behaves significantly different. When the device is operated with a
low driving frequency: $f_{\rm sq}< f_d$, the electron tunnels from
the source to the drain one-by-one in each cycle. Fig.~\ref{fig3}(a)
shows that the electron population changes roughly between zero to
one in each cycle. The corresponding time-dependent left and right
currents flowing into the dot also clearly show that the electron
transfer is dominated by sequential tunnelings. However, with the
higher driving frequency, $f_{\rm sq} > f_d$, the electron
population varies from 0.4 to 0.6 in each cycle, as shown in
Fig.~\ref{fig3}(b). In other words, the driving field oscillates
faster than the response of the electron tunneling such that the
electron only partially tunnels into the dot in each cycle.

\begin{figure}
\includegraphics[width=0.5\columnwidth,angle=0]{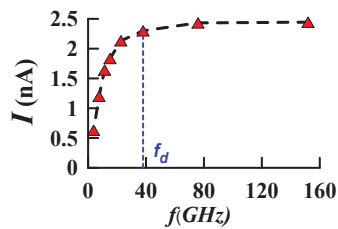}\newline
\caption{(a)The pumping current as a function of the operating
frequency for the square wave modulation.} \label{fig5}
\end{figure}

To have a clear picture how the single-electron turnstile operates
at high frequencies, we plot in Fig.~\ref{fig5} the pumping current
as a function of driving frequency $f_{\rm sq}$ for the square wave
modulation. It shows that when $f_{\rm sq} < f_d/2$, the pumping
current $I = ef_{\rm sq}$ (for $n=1$). This substantiates the
electron transfer one-by-one synchronized with alternating external
gate voltages, as observed in Ref.~\cite{setp1} as well as in other
experimental setup \cite{set4,set5,set6,pumps}. However, for the
operating frequency $ f_{\rm sq} > f_d/2$, the electron transfer
deviates gradually from the above linear relation. When $f_{\rm sq}
> f_d$, the pumping current is saturated and becomes a constant.
As a result, the electron turnstile operation, namely, the electron
transfer one-by-one synchronized with alternating external gate
voltages obeying the relation $I=nef$ works well when $f_{\rm sq} <
f_d/2$.

Same plot is given for the sinusoidal wave modulation, see
Fig.~\ref{fig3}(c)-(d). Different to the square wave modulation,
sinusoidal wave only has a very short period of time that can
totally open the tunneling channels for the source and the drain, as
shown by the time-dependence of $\Gamma_{L,R}(t)$ in
Fig.~\ref{fig2}(b). This makes the electron never have a sufficient
time to fully tunnel into the dot even when the device operates at a
low frequency, $f_{\sin} < f_d$, see the plot in Fig.~\ref{fig3}(c)
where the maximum occupation is about 0.8. Moreover, the average
tunneling rate of the sinusoidal wave modulation is smaller than
that in the square wave modulation for each cycle. Correspondingly
the electron dwell time becomes longer for the sinusoidal wave
modulation. As a result, with a high driving frequency, $f_{\sin}
> f_d$, only a fraction of the electron charge can be tunneled into
the dot in every cycle until the system reaches its steady state, as
shown in Fig.~\ref{fig3}(d). This is indeed a quantum coherence
tunneling, namely the tunneling electron is in a superposition state
between the lead and the dot.

\begin{figure}
\includegraphics[width=1.0\columnwidth,angle=0]{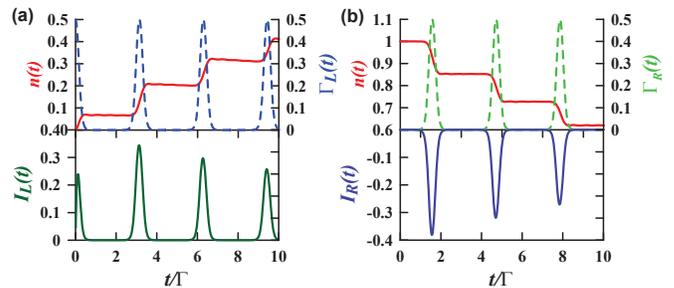}\newline
\caption{The electron population and the transient current when (a)
only the left lead coupling to dot is turned with the sinusoidal
wave modulation. (b) only the right lead coupling to dot is turned
on by the same voltage with initial preparation: $n(t_0)=1$. Here we
take the voltage amplitude $V_{\sin}=1meV$ and the operating
frequency $f_{\sin}$=50GHz . } \label{fig4}
\end{figure}

To make this phenomenon clearer, we may only turn on the left
tunneling (between the source and the dot) but close the right
tunneling (between the dot and the drain). The corresponding results
are plotted in Fig.~\ref{fig4}(a) where the driving frequency,
$f_{\sin}$=50 GHz, is greater than $f_d$. As we can see, in each
cycle the electron transfer shows a plateau of fractional charge. In
other words, in the high driving frequency regime, only fractional
electron rather than the single electron is pumped in each cycle.
The same phenomenon is also seen under a different setup, namely
closing the left tunneling but turning on the right tunneling
channel with the initial occupation $n(t_{0})=1$, the result is
shown in Fig.~\ref{fig4}(b).

In conclusion, our analysis on the transient electron transport show
that the external driving field under different driving frequency
regimes manipulates quite different electron pumping dynamics. When
the external driving frequency is lower than the characteristic
frequency $f_d$ (proportional to the mean average electron tunneling
rate through the dwell time $\tau_d$), the single-electron turnstile
pumping, namely the controllable transfer of electrons one-by-one
synchronized with alternating external gate voltages, works well, as
observed in experiments. Operating the turnstile device at high
frequency with a large pumping current requires an increase of the
effective tunneling rate (shortening the electron response time)
between the central system with its contacts, which is controllable
by the gate voltage profile of the external driving field. As we
have shown, the square wave modulation has a larger $f_d$ than the
sinusoidal wave modulation so that it can operate the
single-electron turnstile with a higher signal frequency for the
same driving field strength. However, the sinusoidal wave modulation
in the high frequency regime leads to the fractional electron
pumping, as a new quantized electron transfer phenomenon. Such
fractional electron pumping phenomenon should be observed in
experiments with the current advances in the nanofabrication
technology, and may also have the potential application in precision
measurement as well as in solid-state quantum computing.

 This work is supported by the National Science
Council of ROC under Contract No. NSC-99-2112-M-006-008-MY3 and
National Center for Theoretical Science.


\begin{thebibliography}{9}




\bibitem{setp1}L. P. Kouwenhoven, A. T.Johnson, N. C. van der Vaart, C. J.
P. M. Harmans, and C. T. Foxon, Phys. Rev. Lett. \textbf{67}, 1626
(1991).

\bibitem{set4} A. Fujiwara, N. M. Zimmerman, Y. Ono, and Y. Takahashi, Appl.
Phys. Lett. \textbf{84}, 1323 (2004); A. Fujiwara, K. Nishiguchi and
Y. Ono, Appl. Phys. Lett. \textbf{92} 042102 (2008).

\bibitem{set5} M. D. Blumenthal, B. Kaestner, L. Li, S. Giblin, T. J. B. M. Janssen,
M. Pepper, D. Anderson, G. Jones and D. A. Ritchie, Nat. Phys.
\textbf{3}, 343 (2007).

\bibitem{set6} J. P. Pekola, J. J. Vartiainen, M. M\"{o}tt\"{o}nen, O.-P. Saira, M.
Meschke, and D. V. Averin, Nat. Phys. \textbf{4}, 120 (2008).

\bibitem{pumps}S. P. Giblin, S. J. Wright, J. D. Fletcher, M. Kataoka, M. Pepper,
T. J. B. M. Janssen, D. A. Ritchie, C. A. Nicoll, D. Anderson, and
G. A. C. Jones, New J. Phys. \textbf{12}, 073013 (2010).




\bibitem{Jin10083013} J. S. Jin, M. W. Y. Tu, W. M. Zhang,
Y. J. Yan, New J. Phys. \textbf{12}, 083013 (2010).

\bibitem{Tu08235311} M. W. Y. Tu and W. M. Zhang, Phys. Rev. B \textbf{78}, 235311 (2008).

\bibitem{Mac06085324} J. Maciejko, J. Wang, and H.
Guo, Phys. Rev. B \textbf{74}, 085324 (2006).

\bibitem{Buttiker96} V. Gasparian, T. Christen, and M. B\"{u}ttiker, Phys. Rev. A \textbf{54}, 4022
(1996).

\bibitem{Guo07}J. Wang, B. Wang, and H. Guo, Phys. Rev. B \textbf{75}, 155336 (2007).











\end{thebibliography}
\end{document}